\documentclass[aps,
 prx,
 reprint,
superscriptaddress,
nofootinbib,
nobibnotes,
amsmath,amssymb,
nolongbibliography,
noeprint,
floatfix
]{revtex4-2}
\usepackage[utf8]{inputenc}
\usepackage{caption}
\usepackage{soul}
\usepackage{ragged2e} % For \justify command (add to preamble)
\usepackage{graphicx}
\usepackage{mathrsfs}  
\usepackage{amsthm}
\usepackage{amssymb} % add new
\usepackage{bm} 
\usepackage{physics}

\usepackage{tikz}
\usepackage{placeins}
\usepackage[pdftex,colorlinks=true,linkcolor=blue,citecolor=blue,urlcolor=blue]{hyperref}
\def\mnras{Mon. not. r. astron. soc.}

\def\apjl{Astrophys. j. lett.}

\def\prd{Phys. rev. d}

\def\nat{Nature}

\begin{document}

\title{Electromagnetic alignment and jet precession around supermassive black holes:\texorpdfstring{\\}{ }Quasi-periodic oscillations in tidal disruption events}

\author{Pau Amaro Seoane}
\affiliation{Universitat Politècnica de València, Spain}
\affiliation{Max Planck Institute for Extraterrestrial Physics, Garching, Germany}
\affiliation{The Kavli Institute for Astronomy and Astrophysics, Beijing, China}

\author{Leif Lui}
\affiliation{Beijing Institute of Mathematical Sciences and Applications, Beijing 101408, China}

\author{Alejandro Torres Orjuela}
\email[Contact author: ]{atorreso@bimsa.cn}
\affiliation{Beijing Institute of Mathematical Sciences and Applications, Beijing 101408, China}

\author{Xian Chen}
\affiliation{Department of Astronomy, School of Physics, Peking University, 100871 Beijing, China}
\affiliation{The Kavli Institute for Astronomy and Astrophysics, Beijing, China}

\begin{abstract}
We evaluate quasi-periodic oscillations and jet formation in tidal disruption events using the covariant formulation of electromagnetic angular-momentum transfer. General-relativistic frame-dragging tears apart misaligned transient accretion flows, forming an isolated inner mini-disk. The accumulation of magnetic flux on the event horizon powers a relativistic jet via the Blandford-Znajek mechanism. Because the magnetic field anchors to the precessing mini-disk, the jet axis rotates, generating geometric modulations in the observed X-ray and radio fluxes. To ensure physical consistency with the force-free magnetosphere required to launch a Blandford-Znajek jet, we model the electromagnetic back-reaction using a split-monopole magnetic field topology. By performing a small-spin expansion of the Noether current density over the event horizon, we derive a closed-form analytical reaction torque exerted by the electromagnetic field on the accretion plasma. We evaluate the resulting kinematics to show that the electromagnetic back-reaction induces a retrograde precession of the mini-disk, coupling with the prograde Lense-Thirring precession to dictate the global oscillation frequency. We formulate explicit predictions for observable transient signals and predict a monotonic attenuation of the peak-to-trough flux ratio as the mini-disk aligns, as well as a specific frequency drift signature characterized by an initial lengthening followed by an asymptotic shortening of the time interval between consecutive flares. We establish an analytical mechanism where magnetic flux depletion stalls alignment, predicting a constant residual modulation amplitude at late times. We formulate a methodology to extract the black hole spin and the magnetic flux density directly from the temporal derivatives of this predicted frequency drift, operating independently of spectral continuum fitting.
\end{abstract}
\maketitle

\section{Introduction}

\label{sec:introduction}

Tidal disruption events (TDEs) generate transient accretion flows that advect external magnetic fields toward supermassive black holes (BHs). The initial stellar debris circularizes into an accretion disk with a random angular momentum orientation relative to the BH spin. High-cadence monitoring of transient events such as at2020ocn isolates quasi-periodic flux modulations operating on timescales of days to weeks \citep{Miniutti2019, PashamEtAl2024, Cao2024}. Resolving the temporal evolution of these quasi-periodic signals requires continuous X-ray or radio monitoring with a sampling cadence strictly shorter than the baseline Lense-Thirring (LT) precession period~\cite{Stone2012}.

Frame dragging induces differential torques that exceed internal viscous communication, tearing the accretion flow and isolating a precessing inner mini-disk \citep{NixonEtAl2012, LiskaEtAl2021}. While stationary vacuum solutions to the Maxwell equations in the Kerr spacetime evaluate exact electromagnetic (EM) torques \citep{Frolov2024}, the formation of a Blandford-Znajek (BZ) jet explicitly requires a highly magnetized, Force-free (FF) plasma to carry the requisite currents ($F_{\mu\nu}J^\nu = 0$)~\citep{BlandfordZnajek1977, Gralla2014}. To satisfy this boundary condition, we model the macroscopic magnetic field using the split-monopole configuration. We apply global angular momentum conservation to the horizon-penetrating Noether current to evaluate the reaction torque on the mini-disk. We model the formation of a relativistic jet from the rotating magnetic field to analyze the quasi-periodic signals. We derive the analytical kinematics of the precessing system, incorporating the EM torque components that generate simultaneous alignment and retrograde precession. We translate the derived kinematic variables into observable domain parameters, mapping the temporal evolution of the flare flux ratio and the observed flare period. We formulate three quantitative predictions for the time-dependent quasi-periodic signals. We predict a monotonic attenuation of the flare amplitude. We predict a late-time residual amplitude floor if magnetic flux depletion halts alignment. We predict a frequency drift exhibiting initial period lengthening followed by asymptotic period shortening. We extract intrinsic system parameters from the temporal derivatives of this predicted frequency drift.

\section{Disk tearing and the precessing mini-disk}
\label{sec:tearing}

The LT effect induces a differential precession frequency $\Omega_{\text{lt}}$ within the misaligned accretion flow. For a BH of dimensionless spin $a$ and mass $M$, the local frequency evaluates to
\begin{equation} \label{eq:lt_omega_local}
    \Omega_{\text{lt}} = \frac{2 a G^2 M^2}{c^3 R^3},
\end{equation}
\noindent where $R$ specifies the radial coordinate. 

\begin{figure}[htbp]
    \centering
    \includegraphics[width=\columnwidth]{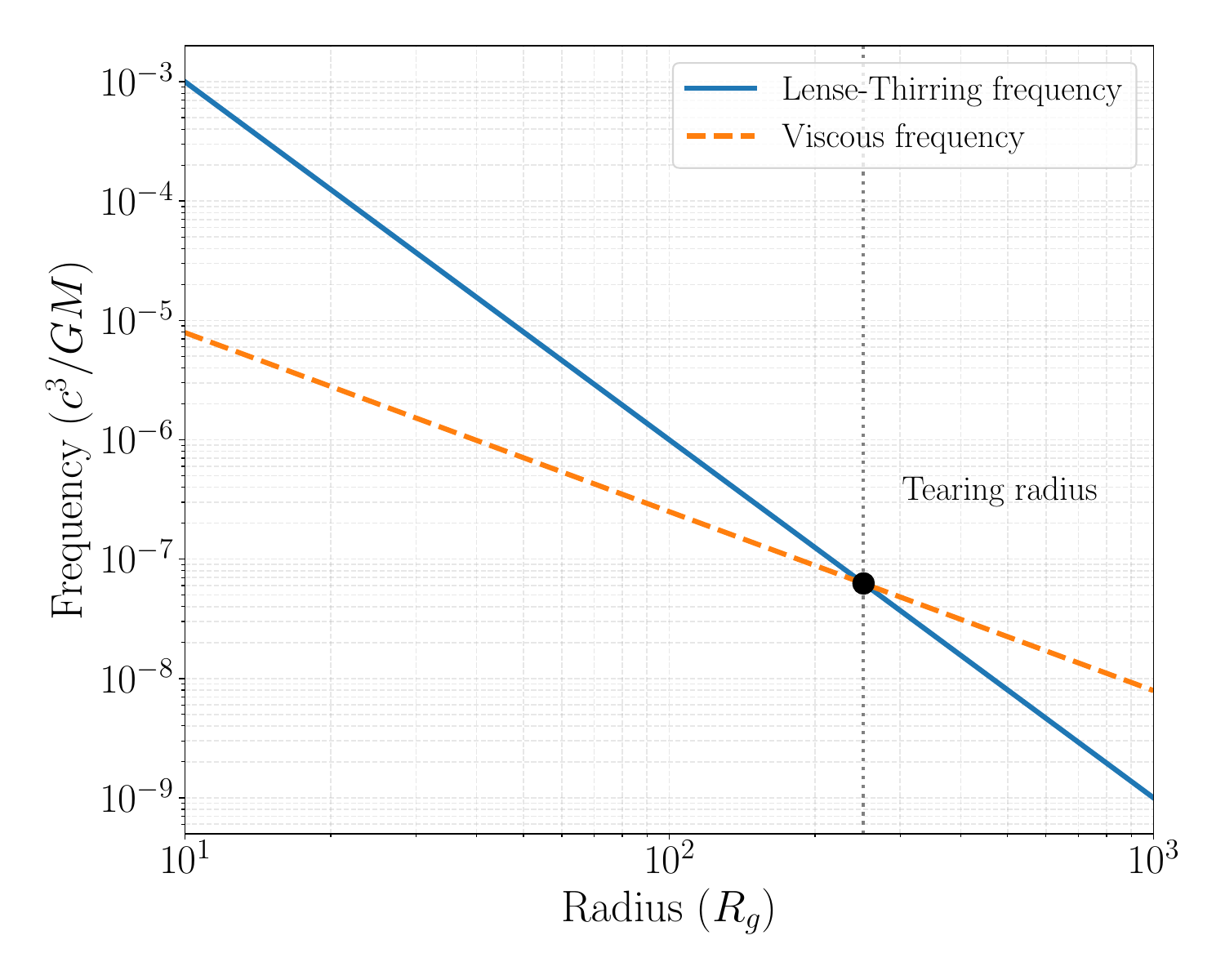}
    \caption{Radial dependence of the characteristic frequencies dictating accretion disk tearing. The differential LT precession frequency evaluates to $10^{-3} c^3 / G M$ at a radius of 10 gravitational radii, decreasing proportionally to the inverse cube of the radius. The internal viscous communication frequency evaluates to $7.91 \times 10^{-6} c^3 / G M$ at 10 gravitational radii. The intersection of these dynamic profiles defines the tearing radius at 251.98 gravitational radii, where both frequencies evaluate to $6.19 \times 10^{-8} c^3 / G M$. At larger radii, such as 636.22 gravitational radii, the viscous frequency exceeds the LT frequency, evaluating to $1.56 \times 10^{-8} c^3 / G M$ compared to $3.88 \times 10^{-9} c^3 / G M$.}
    \label{fig:tearing_frequencies}
\end{figure}

Viscous stresses within the accretion flow oppose this differential rotation. The radial communication of these stresses operates at a local viscous frequency evaluating to $\Omega_{\text{visc}} = \alpha (H/R)^2 \Omega_{\text{k}}$, where $\alpha$ defines the isotropic viscosity parameter, $H/R$ denotes the disk aspect ratio, and $\Omega_{\text{k}}$ specifies the Keplerian orbital frequency. Radiative cooling in TDEs generates geometrically thin accretion flows \citep{LiskaEtAl2021}. We evaluate the communication frequency assuming a viscosity parameter of $\alpha = 0.1$ and a thin-disk aspect ratio of $H/R = 0.05$. Disk tearing occurs at the radius $R_{\text{tear}}$ where the differential LT frequency exceeds this internal viscous communication frequency. Figure \ref{fig:tearing_frequencies} maps the intersection of these two dynamic profiles. The LT frequency evaluates to $10^{-3} c^3 / G M$ at a radius of 10 gravitational radii, exceeding the viscous frequency of $7.91 \times 10^{-6} c^3 / G M$. The frequencies intersect at a tearing radius of 251.98 gravitational radii, where both evaluate to $6.19 \times 10^{-8} c^3 / G M$. This structural separation isolates an inner mini-disk from the misaligned outer accretion flow.

The structural integrity of the mini-disk depends on the efficiency of warp propagation. While the purely viscous communication timescale exceeds the LT precession timescale (causing the outer disk to tear), the intense accretion in the isolated inner flow causes it to geometrically thicken. This thicker structure allows rapid pressure-driven bending waves to dominate internal communication. Because this bending wave timescale evaluates to a value strictly smaller than the local differential precession timescale, the mini-disk resists further tearing and precesses globally as a rigid body~\cite{Papaloizou1983,Fragile2004}. The global precession frequency $\Omega_{\text{prec, global}}$ evaluates as the surface-density weighted average of the local LT frequency over the radial extent of the mini-disk, from the innermost stable circular orbit to the tearing radius. 

The transient accretion flow advects external magnetic flux inward. Orbital shear within the plasma bulk amplifies the internal magnetic field, generating a toroidal topology locally. Because this toroidal field circulates the central mass, the azimuthal components sum to zero during volume integration. The macroscopic magnetic field $\bm{B}_0$ exerting the EM torque and powering the BZ mechanism consists exclusively of the net poloidal flux~\cite{Narayan2003, Tchekhovskoy2011}. The accumulation of this vertical flux establishes a magnetically arrested state where the magnetic pressure balances the radial ram pressure of the accreting gas. The ideal magnetohydrodynamic condition anchors these vertical magnetic field lines to the plasma bulk. Because the mini-disk maintains axisymmetry, the net poloidal vector $\bm{B}_0$ remains perpendicular to the mini-disk geometry, precessing synchronously with the normal vector $\bm{l}$. General-relativistic frame-dragging induces differential torques that govern the structural separation of the transient accretion flow. The tearing radius defines the boundary between the misaligned outer debris and the isolated inner mini-disk. The Kerr metric interacts with the macroscopic poloidal magnetic flux traversing the event horizon, extracting rotational kinetic energy to launch a relativistic jet parallel to the mini-disk normal vector.

\section{Jet formation and geometric modulations}
\label{sec:jet_formation}

The accumulation of magnetic flux in the inner accretion flow fulfills the boundary conditions for the BZ mechanism \citep{BlandfordZnajek1977}. The rotation of the Kerr spacetime twists the magnetic field lines traversing the event horizon, extracting rotational kinetic energy from the BH. The system launches a relativistic jet powered by the outgoing Poynting flux. The EM power extracted scales with the square of the BH spin $a$ and the square of the magnetic flux threading the horizon. Because the magnetic field anchors to the precessing mini-disk, the orientation of the relativistic jet aligns with the time-dependent normal vector $\bm{l}(t)$.

We define $\iota_{\text{obs}}$ as the constant inclination angle of the observer line of sight relative to the BH spin axis $\bm{s}$. As the mini-disk precesses and aligns, the viewing angle $\iota(t)$ between the jet axis and the observer satisfies the spherical trigonometric relation
\begin{equation} \label{eq:viewing_angle}
    \cos\iota(t) = \cos\iota_{\text{obs}} \cos\theta(t) + \sin\iota_{\text{obs}} \sin\theta(t) \cos\Phi(t),
\end{equation}
\noindent where $\Phi(t) = \int \Omega_{\text{prec}}(t) dt$ defines the integrated precession phase.

This geometric modulation dictates the observed flux through relativistic Doppler beaming. The Doppler factor $\delta(t)$ evaluates to
\begin{equation} \label{eq:doppler_factor}
    \delta(t) = \frac{1}{\Gamma [1 - \beta \cos \iota(t)]},
\end{equation}
\noindent where $\Gamma$ denotes the bulk Lorentz factor of the jet and $\beta$ denotes the jet velocity normalized by the speed of light.

The observed monochromatic flux scales proportionally to $\delta(t)^{3+\alpha_{\nu}}$, where $\alpha_{\nu}$ defines the spectral index of the emission. Equation~(\ref{eq:viewing_angle}) couples the Doppler beaming mechanism to the EM back-reaction. The periodic variation of the azimuthal phase $\Phi(t)$ increases the emitted flux when the jet rotates toward the observer and decreases the flux when the jet rotates away. Concurrently, as the alignment torque drives the misalignment angle $\theta(t)$ toward the equatorial plane, the amplitude of the second term in Eq.~(\ref{eq:viewing_angle}) decays. This kinematic coupling dictates that the geometric alignment of the mini-disk reduces the variance of the viewing angle, driving the attenuation of the quasi-periodic flare amplitudes~\cite{Middleton2025}.

\section{Covariant electromagnetic precession and alignment}
\label{sec:em_kinematics}

The extraction of rotational energy and angular momentum via the BZ mechanism dictates the back-reaction on the mini-disk. 
We evaluate the axial spin-down component of the EM torque $\mathcal{T}^z_{\text{bh}}$ acting on the rotating BH by integrating the radial flux of azimuthal angular momentum over the event horizon $\mathcal{H}$,~\cite{Lasota2014, Gralla2014}
\begin{equation} \label{eq:torque_integral}
    \mathcal{T}^z_{\text{bh}} = \int_{\mathcal{H}} T^r_{\ \phi} \sqrt{-g} \, d\theta d\phi,
\end{equation}
\noindent where $T^{\mu}_{\ \nu}$ specifies the energy-momentum tensor of the FF split-monopole magnetosphere. The full three-dimensional torque vector $\bm{\mathcal{T}}_{\text{bh}}$ is formally constructed by projecting the energy-momentum tensor onto the respective spatial rotational Killing vectors. Integrating the divergence of the total angular momentum current over the spacetime volume gives us the global conservation law. Because the outgoing BZ jet channels the massive parallel spin-down angular momentum to infinity ($\dot{\bm{J}}_{\text{jet}}$), the mini-disk specifically absorbs only the transverse (perpendicular) back-reaction components of the BH torque. The rate of change of the mini-disk angular momentum $\bm{J}_{\text{mini}}$ balances this transverse component, yielding $\bm{\mathcal{T}}_{\text{mini}} = - \bm{\mathcal{T}}_{\text{bh},\perp}$.

Because evaluating the full vector torque integrals at the horizon limit $r_H = M + M\sqrt{1-a^2}$ is analytically intractable for an arbitrary dimensionless spin $a$, we expand the integrands as a Taylor series. Expanding up to second order $\mathcal{O}(a^2)$ captures both the dominant $\mathcal{O}(a)$ alignment effect and the sub-leading $\mathcal{O}(a^2)$ BZ precession scaling, yielding closed-form analytical expressions for the transverse torques~\cite{Tchekhovskoy2010}.

The resulting expanded torque separates into an alignment component $\bm{\mathcal{T}}_1$ operating orthogonal to the mini-disk and a precession component $\bm{\mathcal{T}}_2$. The alignment torque component operates perpendicular to the mini-disk angular momentum vector. The macroscopic magnetic field anchors to the mini-disk, establishing $\bm{B}_0 = B_0 \bm{l}$. Substituting this condition into the alignment torque equation yields
\begin{equation} \label{eq:t1_torque_substituted}
    \bm{\mathcal{T}}_1 = -\mathcal{C}_{\text{align}} M B_0^2 \bm{l} \times (\bm{J}_{\text{bh}} \times \bm{l}).
\end{equation}
\noindent The outer cross product dictates that the resulting torque vector operates orthogonal to $\bm{l}$. Evaluating the dot product confirms this orthogonality. The vector triple product expansion produces
\begin{equation} \label{eq:vector_triple_product}
    \bm{l} \times (\bm{J}_{\text{bh}} \times \bm{l}) = \bm{J}_{\text{bh}}(\bm{l} \cdot \bm{l}) - \bm{l}(\bm{l} \cdot \bm{J}_{\text{bh}}).
\end{equation}
\noindent Because $\bm{l}$ constitutes a unit vector, the dot product $\bm{l} \cdot \bm{l}$ evaluates to 1. Taking the dot product of the expanded term with $\bm{l}$ gives us
\begin{equation} \label{eq:dot_product_expansion}
    \bm{l} \cdot [\bm{J}_{\text{bh}} - \bm{l}(\bm{l} \cdot \bm{J}_{\text{bh}})] = (\bm{l} \cdot \bm{J}_{\text{bh}}) - (\bm{l} \cdot \bm{J}_{\text{bh}}),
\end{equation}
\noindent which evaluates to zero. This orthogonality dictates that the EM torque alters the geometric orientation of the mini-disk without modifying the scalar magnitude of its angular momentum.

We define $\bm{s} = \bm{J}_{\text{bh}} / J_{\text{bh}}$ as the unit vector of the BH spin and $\bm{l} = \bm{J}_{\text{mini}} / J_{\text{mini}}$ as the unit vector of the mini-disk angular momentum. The anchored magnetic field establishes $\bm{B}_0 = B_0 \bm{l}$. We define $\theta$ as the misalignment angle, establishing $\cos\theta = \bm{s} \cdot \bm{l}$. The total reaction torque and the LT effect dictate the temporal evolution of the unit vector $\bm{l}$, producing the kinematic equation
\begin{equation} \label{eq:l_dot}
    \frac{\dd \bm{l}}{\dd t} = \Omega_{\text{lt}} (\bm{s} \times \bm{l}) + \frac{1}{J_{\text{mini}}} \left( \bm{\mathcal{T}}_{\text{align}} + \bm{\mathcal{T}}_{\text{prec}} \right),
\end{equation}
\noindent where $\bm{\mathcal{T}}_{\text{align}} = -\bm{\mathcal{T}}_1$ and $\bm{\mathcal{T}}_{\text{prec}} = -\bm{\mathcal{T}}_2$. Substituting the vector relations into the components leads to
\begin{equation} \label{eq:l_dot_expanded}
    \frac{\dd\bm{l}}{\dd t} = \left( \Omega_{\text{lt}} - \Omega_{\text{em,0}} \cos\theta \right) (\bm{s} \times \bm{l}) + \frac{1}{\tau_{\text{align,0}}} \left[ \bm{s} - (\bm{s} \cdot \bm{l})\bm{l} \right],
\end{equation}
\noindent where the initial maximum EM precession amplitude $\Omega_{\text{em,0}}$ and the initial alignment timescale $\tau_{\text{align,0}}$ evaluate to
\begin{equation} \label{eq:omega_em}
    \Omega_{\text{em,0}} = \mathcal{C}_{\text{prec}} \frac{B_0^2 J_{\text{bh}}^2}{M J_{\text{mini}}}, \quad \frac{1}{\tau_{\text{align,0}}} = \mathcal{C}_{\text{align}} \frac{M B_0^2 J_{\text{bh}}}{J_{\text{mini}}}.
\end{equation}
\noindent The dimensionless coefficients $\mathcal{C}_{\text{prec}}$ and $\mathcal{C}_{\text{align}}$ parameterize the scaling of the torques and are formally determined by the $\mathcal{O}(a^2)$ expansion of the FF magnetosphere. The total precession frequency evaluates to $\Omega_{\text{prec}} = \Omega_{\text{lt}} - \Omega_{\text{em,0}} \cos\theta$. The negative sign dictates that the EM torque induces a retrograde precession relative to the BH spin vector. Concurrently, taking the dot product of Eq.~(\ref{eq:l_dot_expanded}) with the stationary vector $\bm{s}$ extracts the differential equation for the misalignment angle, producing
\begin{equation} \label{eq:theta_dot}
    \frac{\dd \theta}{\dd t} = - \frac{1}{\tau_{\text{align,0}}} \sin\theta.
\end{equation}
\noindent The EM field back-reaction simultaneously precesses the mini-disk and drives it toward the equatorial plane of the BH.

\section{Time-dependent fallback kinematics}
\label{sec:time_dependent}

During a tidal disruption event, the mass fallback rate decays according to the theoretical profile $\dot{M}(t) = \dot{M}_0 (1 + t/t_{\text{fb}})^{-5/3}$, where $t_{\text{fb}}$ denotes the fallback timescale~\cite{Phinney1989, Rees1988}. The magnetic energy density of the arrested disk scales proportionally to the mass accretion rate, imposing the temporal decay $B_0^2(t) = B_{\text{max}}^2 (1 + t/t_{\text{fb}})^{-5/3}$. Substituting this condition into the alignment differential equation formulated in Eq.~(\ref{eq:theta_dot}) yields a separable equation for the misalignment angle, evaluating to~
\begin{equation} \label{eq:theta_decay}
    \frac{\dd\theta}{\dd t} = - \frac{1}{\tau_{\text{align,0}}} \left( 1 + \frac{t}{t_{\text{fb}}} \right)^{-5/3} \sin\theta,
\end{equation}
\noindent where $\tau_{\text{align,0}}$ specifies the initial alignment timescale. Rearranging and integrating this equation from the initial state produces the analytical time-dependent alignment trajectory
\begin{equation} \label{eq:theta_analytic}
\begin{split}
    \theta(t) =  2 \arctan\left[ \tan\left( \frac{\theta_0}{2} \right) \exp\left( - \frac{3 t_{\text{fb}}}{2 \tau_{\text{align,0}}}\right.\right. \\
     \left.\left. \times\left[ 1- \left( 1 + \frac{t}{t_{\text{fb}}} \right)^{-2/3} \right] \right) \right].
\end{split}
\end{equation}
\noindent As the time parameter approaches infinity, the temporal term approaches zero. The exponential argument asymptotes to $-3 t_{\text{fb}} / (2 \tau_{\text{align,0}})$. Consequently, the misalignment angle converges to a non-zero constant. This limit establishes an alignment freeze-out mechanism. The decay of the mass fallback rate depletes the magnetic energy density required to enforce complete alignment.

The retrograde EM precession couples with the decaying mass supply, dictating the time-dependent global precession period of the transient signal. Substituting the fallback condition into the kinematic equations gives the analytical period
\begin{equation} \label{eq:t_prec_analytic}
    T_{\text{prec}}(t) = 2\pi \left[ \Omega_{\text{lt}} - \Omega_{\text{em,0}} \left( 1 + \frac{t}{t_{\text{fb}}} \right)^{-5/3} \cos\theta(t) \right]^{-1}.
\end{equation}

\section{Frequency drift extraction of system parameters}
\label{sec:frequency_drift}

The frequency drift provides a methodology to extract the BH spin. Time-series analysis over the duration of the transient event isolates the first derivative of the frequency, $d\Omega_{\text{prec}}/dt$. In practice, extracting this derivative requires dividing the continuous light curve into discrete temporal segments. Techniques such as dynamic power spectral density estimation or localized epoch-folding generate a sequence of discrete frequency measurements, which are subsequently differentiated to yield $d\Omega_{\text{prec}}/dt$.

We evaluate the analytical derivative of the global precession frequency to isolate the system parameters. The frequency evaluates to $\Omega_{\text{prec}}(t) = \Omega_{\text{lt}} - \Omega_{\text{em,0}} ( 1 + t/t_{\text{fb}} )^{-5/3} \cos\theta(t)$. Differentiating the frequency with respect to time, we have that

\begin{equation} \label{eq:omega_prec_derivative}
    \frac{\dd \Omega_{\text{prec}}}{\dd t} = \frac{\dd\Omega_{\text{lt}}}{\dd t} - \frac{\dd}{\dd t} \left[ \Omega_{\text{em,0}} \left( 1 + \frac{t}{t_{\text{fb}}} \right)^{-5/3} \cos\theta(t) \right].
\end{equation}
\noindent The LT frequency depends on the BH spin and the tearing radius. Assuming the tearing radius remains constant, the temporal derivative of the LT frequency evaluates to zero. Applying the product rule to the second term produces
\begin{align} \label{eq:product_rule}
    \frac{\dd\Omega_{\text{prec}}}{\dd t} = \Omega_{\text{em,0}} & \left[ \frac{5}{3 t_{\text{fb}}} \left( 1 + \frac{t}{t_{\text{fb}}} \right)^{-8/3} \cos\theta(t) + \right. \nonumber \\
    & \left. \left( 1 + \frac{t}{t_{\text{fb}}} \right)^{-5/3} \sin\theta(t) \frac{\dd\theta}{\dd t} \right].
\end{align}
\noindent We substitute the differential equation for the misalignment angle into Eq.~(\ref{eq:product_rule}). This substitution gives

\begin{align} \label{eq:domega_prec_dt_expanded}
    \frac{\dd\Omega_{\text{prec}}}{\dd t} = &\Omega_{\text{em,0}} \left( 1 + \frac{t}{t_{\text{fb}}} \right)^{-5/3}  \left[ \frac{5}{3 t_{\text{fb}}} \left( 1 + \frac{t}{t_{\text{fb}}} \right)^{-1} \right. \nonumber \\
    & \times\cos\theta(t) - \left. \frac{1}{\tau_{\text{align,0}}} \left( 1 + \frac{t}{t_{\text{fb}}} \right)^{-5/3} \sin^2\theta(t) \right].
\end{align}

\noindent The measurement of the initial frequency derivative isolates the initial EM precession amplitude. Evaluating Eq.~(\ref{eq:domega_prec_dt_expanded}) at an initial time provides
\begin{equation} \label{eq:domega_prec_dt_zero}
    \left.\frac{\dd \Omega_{\text{prec}}}{\dd t} \right|_{t=0} = \Omega_{\text{em,0}} \left( \frac{5}{3 t_{\text{fb}}} \cos\theta_{0} - \frac{1}{\tau_{\text{align,0}}} \sin^2\theta_0 \right).
\end{equation}
\noindent The peak-to-trough amplitude decay of the modulations over the observation window constrains the initial inclination and the alignment timescale. The broadband light curve of the transient event constrains the fallback timescale. Substituting these measured values into Eq.~(\ref{eq:domega_prec_dt_zero}) gives the value of the EM precession amplitude.

\begin{figure*}[htbp]
    \centering
    \includegraphics[width=\textwidth]{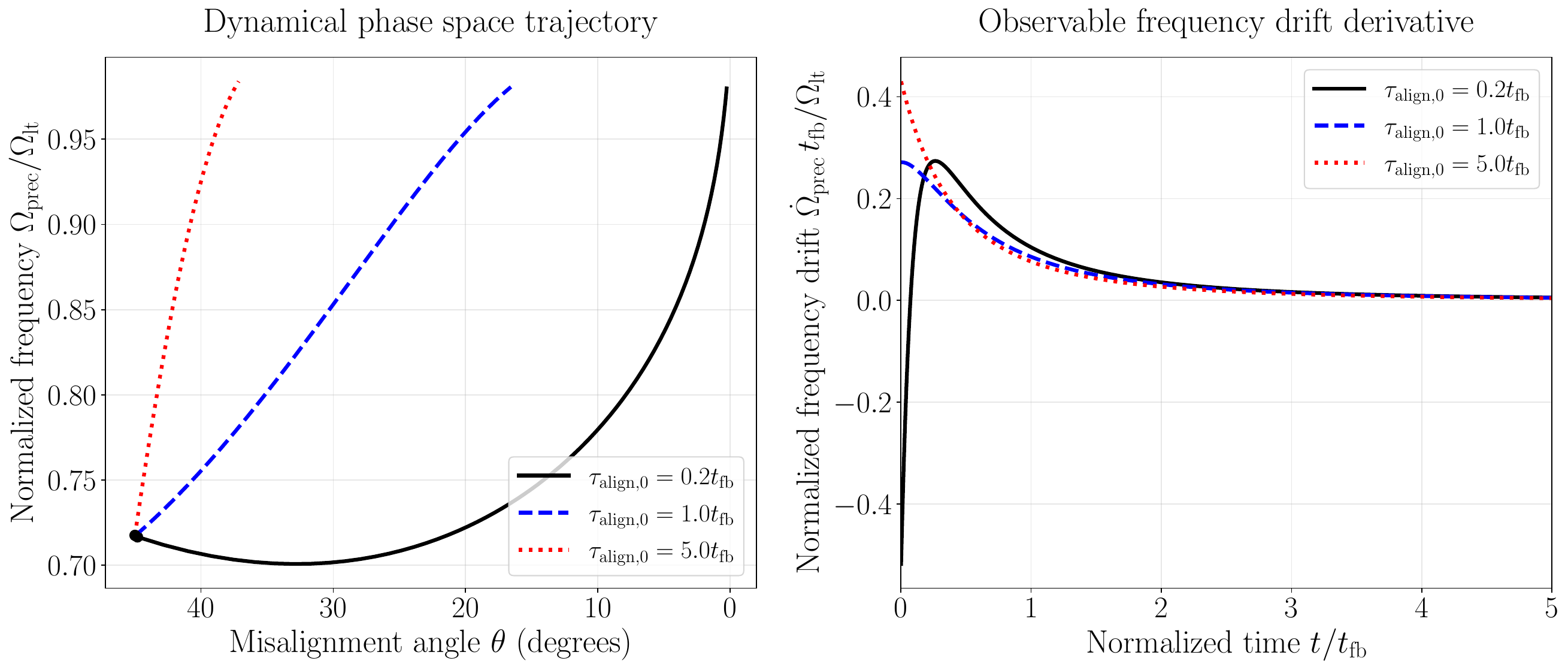}
    \caption{Extraction parameters for the precessing mini-disk model, tracking scenarios with different alignment speeds relative to the material fallback time (assuming an initial EM precession ratio of 0.4 and a 45-degree tilt). The left panel maps the phase space trajectories of the normalized global precession frequency against the shrinking misalignment angle, with time progressing from right to left as the disk aligns from its starting coordinate. The right panel illustrates how the frequency's rate of change (the drift derivative) evolves. In the fast-alignment scenario, this derivative begins moderately negative, quickly shifts to positive very early in the evolution, and ultimately flattens out toward zero as the mass fallback rate dwindles.}
    \label{fig:frequency_drift}
\end{figure*}

From Eq.~(\ref{eq:omega_em}), the ratio of the initial EM precession amplitude to the inverse initial alignment timescale defines a characteristic coupling constant $\kappa = \mathcal{C}_{\text{prec}} / \mathcal{C}_{\text{align}}$, evaluating to
\begin{equation} \label{eq:ratio_omega_tau}
    \Omega_{\text{em,0}} \tau_{\text{align,0}} = \kappa a.
\end{equation}
\noindent Rearranging this relation yields a deterministic measurement of the dimensionless BH spin $a$, producing
\begin{equation} \label{eq:spin_measurement}
    a = \frac{1}{\kappa} \Omega_{\text{em,0}} \tau_{\text{align,0}}.
\end{equation}

\noindent The initial global precession frequency evaluates to $\Omega_{\text{prec}}(0) = \Omega_{\text{lt}} - \Omega_{\text{em,0}} \cos\theta_0$. The measurement of the initial frequency and the extraction of the EM precession amplitude define the LT frequency. The host galaxy's stellar velocity dispersion constrains the BH mass. Substituting the derived spin and the LT frequency back into Eq.~(\ref{eq:lt_omega_local}) determines the tearing radius. 

The derivation of the BH spin operates independently of the mass of the mini-disk because the angular momentum parameter cancels in the product ratio. However, extracting the absolute initial magnetic field requires an explicit parameterization of the mini-disk angular momentum to break the geometric degeneracy. Evaluating the absolute magnetic flux requires prior constraints on the mass fallback and accretion efficiency.

Figure \ref{fig:frequency_drift} translates the analytical kinematics into the observable parameter space. The phase space trajectory connects the unobservable geometric coordinate to the observable frequency domain. The temporal derivative of the global precession frequency constitutes the primary measurement required to invert the model. The magnitude of the initial frequency derivative determines the EM precession amplitude, operating independently of broadband spectral continuum assumptions.

\section{Predictions for transient signals}
\label{sec:predictions}

\begin{figure*}[htbp]
    \centering
    \includegraphics[width=\textwidth]{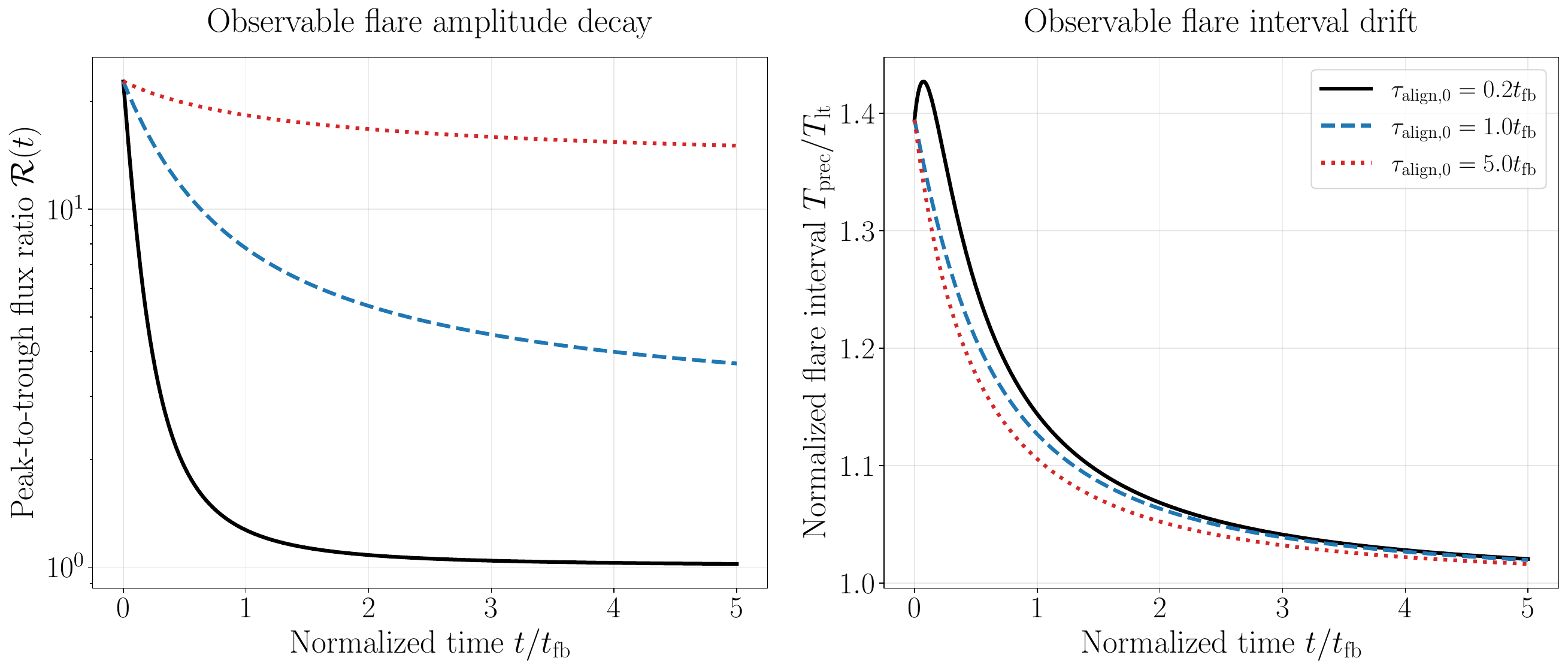}
    \caption{Calculated temporal evolution of observable transient signals generated by a precessing mini-disk and relativistic jet, highlighting scenarios of fast, moderate, and slow (or stalled) alignment relative to the material fallback time. The left panel shows the decay of the peak-to-trough flux ratio from a common pronounced baseline, assuming a 45-degree initial tilt, a 60-degree viewing angle, a jet velocity of 0.5$c$, and a spectral index of 1. In the fast alignment scenario, the signal reduces by a factor of roughly 18 by time 1 and nearly disappears by time 5; moderate alignment leads to a reduction by a factor of about 3 by time 1 and roughly 6 by time 5; whereas slow alignment, which stalls due to magnetic flux depletion, results in only a 20\% drop by time 1 and a roughly one-third reduction by time 5. The right panel displays the corresponding evolution of the normalized global precession period, assuming an initial EM precession ratio of 0.4. From a shared elevated baseline, the fast alignment period briefly increases before dropping, while overall, as the mass fallback rate decreases, all configurations asymptotically decay toward a baseline of unity, converging by time 5 regardless of their initial alignment speed.}
    \label{fig:observable_signals}
\end{figure*}

We translate the analytical kinematics into measurable observables for high-cadence light curves. Observers extract two primary time-dependent quantities during the transient event. These quantities comprise the peak-to-trough flux ratio of the quasi-periodic flares and the temporal interval between consecutive flare peaks. Figure \ref{fig:observable_signals} displays the numerical evaluation of the temporal trajectories dictating these observables.

We define the observable peak-to-trough flux ratio $\mathcal{R}(t)$ as the quotient of the maximum and minimum fluxes measured during a single precession cycle. Evaluating the viewing angle from Eq.~(\ref{eq:viewing_angle}) at the azimuthal phases $\Phi = 0$ and $\Phi = \pi$ gives the maximum and minimum Doppler factors. Assuming the observer inclination $\iota_{\text{obs}}$ exceeds the misalignment angle $\theta(t)$, the observed flux ratio evaluates to
\begin{equation} \label{eq:flux_ratio}
    \mathcal{R}(t) = \left[ \frac{1 - \beta \cos(\iota_{\text{obs}} + \theta(t))}{1 - \beta \cos(\iota_{\text{obs}} - \theta(t))} \right]^{3+\alpha_{\nu}}.
\end{equation}

\noindent As the misalignment angle decays toward zero, the numerator approaches the denominator. The left panel of Figure \ref{fig:observable_signals} demonstrates that the observed flux ratio $\mathcal{R}(t)$ decays monotonically toward unity. High-cadence monitoring measures this attenuation to track the geometric alignment of the mini-disk. 

The alignment freeze-out mechanism produces a measurable signature in the late-time light curve. If the initial alignment timescale exceeds the fallback timescale, the depletion of the accreting magnetic flux prevents complete geometric alignment. For an alignment timescale of $5.0 t_{\text{fb}}$, the flux ratio stalls at a value greater than unity. In this configuration, the observer measures a constant residual modulation amplitude at late times. This residual amplitude isolates the initial magnetic flux density.

The temporal interval between consecutive X-ray flares measures the global precession period $T_{\text{prec}}(t)$. Observationally, this quantity corresponds to the time delay between adjacent flux maxima. Tracking the centroid positions of these maxima across the observation window isolates the geometric precession timescale from the stochastic variability of the continuum. The right panel of Figure \ref{fig:observable_signals} displays the temporal signature of the EM back-reaction. For systems possessing an alignment timescale of $0.2 t_{\text{fb}}$, the retrograde EM torque determines the kinematic evolution at early times. As the misalignment angle decays, the cosine multiplier in Eq.~(\ref{eq:t_prec_analytic}) increases, amplifying the retrograde precession. The observed time between flares increases, reaching a maximum period. Subsequently, the decay of the mass fallback rate suppresses the magnetic energy density required to sustain the retrograde torque. The time between consecutive flares then decreases, asymptotically approaching the constant LT period. The measurement of this period lengthening followed by period shortening isolates the EM back-reaction from purely general relativistic precession.

\section{Conclusions}
\label{sec:conclusions}

We evaluate the quasi-periodic oscillations and jet formation in TDEs using the kinematics of accretion disk tearing and covariant EM angular momentum transfer. General relativistic frame dragging tears the misaligned transient accretion flow, isolating an inner mini-disk that undergoes rigid-body precession. The magnetic field anchored to the mini-disk extracts rotational energy from the BH via the BZ mechanism, launching a relativistic jet. The rotation of the jet axis produces geometric modulations in the observed X-ray and radio fluxes.

We utilize a small-spin expansion of the FF split-monopole magnetosphere to evaluate the reaction of the EM field on the accretion plasma. The back-reaction induces a retrograde precession that couples with the prograde LT effect to determine the global oscillation period. Concurrently, the EM alignment torque drives the mini-disk toward the equatorial plane. We translate this coupled kinematic model into predictions for specific observable quantities. We formulate three explicit predictions for transient light curves. First, we predict a monotonic decay of the peak-to-trough flux ratio driven by geometric alignment. Second, we predict a residual modulation amplitude at late times generated by magnetic flux depletion stalling the alignment process. Third, we predict a specific timing signature where the interval between consecutive flares lengthens initially before shortening asymptotically toward the LT period. We formulate a methodology utilizing the temporal derivatives of these predicted signals to extract the BH spin and the magnetic flux density directly from the high-cadence light curves.

\section*{Acknowledgments}

We acknowledge support from the National Foreign Expert Program (H). ATO acknowledges support from the National Science Foundation of China (No. W2533010). ATO and LL were supported by the Beijing Natural Science Foundation (No. IS25014). 

\bibliographystyle{aasjournalv7}
%\bibliography{biblio}

\end{document}